\documentclass[twocolumn]{aastex63}

\usepackage{amsmath}
\usepackage{threeparttable}

\usepackage{CJKutf8}

\submitjournal{ApJ}
\shortauthors{Liu et al.}

\usepackage{lineno}
\begin{document}
	
	\begin{CJK*}{UTF8}{gbsn}

		\title{The Statistical Polarization Properties of Coherent Curvature Radiation by Bunches: Application to Fast Radio Burst Repeaters}		
		
		\author{Ze-Nan Liu}
		\affiliation{School of Astronomy and Space Science, Nanjing University, Nanjing 210023, China}
		\affiliation{Key Laboratory of Modern Astronomy and Astrophysics (Nanjing University), Ministry of Education, Nanjing 210023, China}
		
		\author{Wei-Yang Wang}
		\affiliation{School of Physics and State Key Laboratory of Nuclear Physics and Technology, Peking University, Beijing 100871, P.R.China}
		\affiliation{Kavli Institute for Astronomy and Astrophysics, Peking University, Beijing 100871, P.R.China}
		
		\author{Yuan-Pei Yang}
		\affiliation{South-Western Institute for Astronomy Research, Yunnan University, Kunming 650500, China}
		\affiliation{Purple Mountain Observatory, Chinese Academy of Sciences, Nanjing 210023, Peopleʼs Republic of China}

		\author{Zi-Gao Dai}
		\affiliation{Department of Astronomy, University of Science and Technology of China, Hefei 230026, China; daizg@ustc.edu.cn}
		\affiliation{School of Astronomy and Space Science, Nanjing University, Nanjing 210023, China}

		\begin{abstract}		
			Fast radio bursts (FRBs) are extragalactic radio transients with millisecond duration and extremely high brightness temperature. Very recently, some highly circularly polarized bursts were found in a repeater, FRB 20201124A. The significant circular polarization might be produced by coherent curvature radiation by bunches with the line of sight (LOS) deviating from the bunch central trajectories. In this work, we carry out simulations to study the statistical properties of burst polarization within the framework of coherent curvature radiation by charged bunches in the neutron star magnetosphere for repeating FRBs. The flux is almost constant within the opening angle of the bunch.  However, when the LOS derivates from the bunch opening angle, the larger the derivation, the larger the circular polarization but the lower the flux.  We investigate the statistical distribution of circular polarization and flux of radio bursts from an FRB repeater, and find that most of the bursts with high circular polarization have a relatively low flux. Besides, we find that most of the depolarization degrees of bursts have a small variation in a wide frequency band. Furthermore, we simulate the polarization angle (PA) evolution and find that most bursts show a flat PA evolution within the burst phases, and some bursts present a swing of PA.
			
		\end{abstract}
		
		
		\section{introduction}
	\end{CJK*}	
	Fast radio bursts (FRBs) are millisecond duration radio transients \citep{Lorimer2007,Keane2012,Thornton2013},  and their physical origins are still mysterious. So far, there are hundreds of FRBs that have been detected, and a small fraction of them show repeating behaviors\footnote{A Transient Name Server system for newly reported FRBs \citep[https://www.wis-tns.org,][]{Petroff2020}.}. Recently, \cite{Bochenek2020} and \cite{CHIME2020b} reported that a bright FRB-like burst from a Galactic magnetar SGR J1935+2154, suggesting that magnetars are the very likely physical origins of FRBs. Many theoretical models have been proposed to explain FRBs \citep[see details from ][]{Platts2019,ZB2020Natur,XWD2020,Lyubarsky2021}. According to the position of radiation region, they are classified into two categories: close-in models, i.e., inside the neutron star magnetosphere \citep[e.g.,][]{Kumar2017,Yang2018,Yang2021,Lu2020,Wang2019,Wang2020} and far-away models, i.e., outside the neutron star magnetosphere \citep[e.g.,][]{Lyubarsky2014,Beloborodov2019,Metzger2019,Margalit2020b}.

	Polarization is  an important probe to study  the emission mechanism and is related to the geometry of the emission region. The polarization features of FRBs appear diversity. Most FRBs have strong linear polarization (LP) fractions near 100$\%$ and have a flat PA across each pulse \citep{ DS2021,Nimmo2021,CHIME2019,Gajjar2018,Michilli2018}. However, some sources (e.g., FRB 180301 and FRB 181112) show partial linear or circular polarization and PAs vary remarkably with time \citep{Luo2020,Cho2020}. Notably, an ‘S’ or ‘inverse S’ shape pattern generally can be observed in the radio pulsars \citep{Lorimer2012}. The profiles have been predicted by the rotating vector model \citep{RVM1969}. A significant circular polarization (CP) was observed in an FRB repeater, FRB 20201124A \citep{jiang2022,Hilmarsson2021,Kumar2021,XuH2021}. A burst with a significant CP fraction can be as high as 47$\%$ \citep{Kumar2021}. Surprisingly, the CP fraction can be up to 75$\%$ found in a good fraction of bursts \citep{XuH2021}.  Very recently, \cite{jiang2022} reported  most of the bursts with low CP degree and a small part of bursts have CP fraction higher than $70\%$. The highly CP fraction may be caused by an intrinsic radiation mechanism \citep{WangJiang2021,WangYang2021,Tong2022} or propagation effects \citep{Beniamini2022}.

	In this paper, we mainly investigate the statistical features of polarization of radio bursts from an FRB repeater within the framework of coherent curvature radiation for repeating FRBs, as proposed in some models \citep[e.g.,][]{Kumar2017,Yang2018,Lu2020,Wang2019,Wang2020}. The paper is organized as follows.
	We first discuss the coherent curvature emission by charged bunches in the neutron star magnetosphere in Section \ref{sec2}. The statistical properties of polarization of radio bursts and the simulation results are shown in Section \ref{sec3}. The results are discussed and summarized in Section \ref{sec4}. The convention $Q_x = Q/10^x$ in cgs units is used throughout this paper.

	\section{FRBs from Coherent Curvature Emission by Charges in the Neutron Star Magnetosphere} \label{sec2}

	Coherent curvature emission is emitted when the relativistic particles move along the curved magnetic field lines.
	Some models invoking coherent curvature emission by charged bunches can explain the radio emission of pulsars
	\citep{Ruderman1975,Sturrock1975,Elsaesser1976,Cheng1977,Melikidze2000,Gil 2004,Gangadhara2021} and FRBs \citep{Katz2014,Kumar2017,Yang2018,Ghisellini 2018,LuKumar2018,Katz2018,WangLai2020,Wang2020,Cooper 2021,WangYang2021}.
	When the size of the charged bunch is smaller than the half wavelength, electromagnetic waves are coherently enhanced remarkably. If the mean space among bunches is much larger than $\lambda/2$, particles among bunches are incoherent \citep{Yang2018,WangYang2021}. The outflow from the neutron star polar cap is probably unsteady. Because of the interaction between near-surface parallel electric field and charged particles from the surface, the neutron star inner gap would generate inhomogeneous sparks \citep{Ruderman1975,Zhang1996,Gil2000}.
	These sparks of electron-positron pairs would form bunches in the magnetosphere of a neutron star by two-stream instability \citep{Usov1987,Asseo1998,Melikidze2000}. In this section, we mainly consider a three-dimensional bunch case to study the polarization properties of radio bursts from an FRB repeater.

	Consider an electron that moves along a trajectory $\boldsymbol{r}(t)$ and the unit vector of the line of sight (LOS) is denoted by $\boldsymbol{n}$.  We assume that a three-dimensional bunch is uniformly distributed in all directions of $\left(\boldsymbol{e}_{r}, \boldsymbol{e}_{\theta}, \boldsymbol{e}_{\phi}\right)$.
	A coherent summation of the amplitudes should replace the single amplitude for a three-dimensional  bunch case.  The total energy radiated per unit solid angle per unit frequency interval for the moving charges in the magnetosphere can be written as \citep{WangYang2021}
	
	\begin{align}
	&\frac{d^{2} W}{d \omega d \Omega}=\frac{e^{2} \omega^{2}}{4 \pi^{2} c}\notag \\ &\times\left|\int_{-\infty}^{+\infty} \sum_{i}^{N_{s}} \sum_{j}^{N_{\theta}} \sum_{k}^{N_{\phi}}-\boldsymbol{\beta}_{e \perp, i j k} \mathrm{e}^{i \omega\left[t-\boldsymbol{n} \cdot \boldsymbol{r}_{i j k}(t) / c\right]} d t\right|^{2},\label{eq1}
	\end{align}where $\omega$ is the observed angular frequency, $t$ is the retarded time, $c$ is the speed of light, and $e$ is the elementary charge. \(N_{s}, N_{\theta}, N_{\phi}\) are the maximum numbers for the identifiers of charge \(i, j, k\) and the total number of positron is \(N=N_{s} N_{\theta} N_{\phi}\). $\boldsymbol{\beta}_{e \perp}$ is defined as the component of $\boldsymbol{\beta}_{e}$ in the plane that is vertical to the LOS:  $
	\boldsymbol{\beta}_{e \perp}=-\boldsymbol{n} \times\left(\boldsymbol{n} \times \boldsymbol{\beta}_{e}\right)
	$, where $
	\boldsymbol{\beta}_{e}= v_e/ c
	$, $v_e$ is the electron velocity.  We define $\chi$ as the angle between the electron velocity direction and the $x$-axis at $t$ = 0 for one trajectory, and  $\varphi$ is defined as the angle between the LOS and the trajectory plane.  The total amplitude of $i$th bunch projection at the plane perpendicular to the LOS is then given by \citep{WangYang2021}
	
	\begin{align}
	&A_{\|, i}\simeq \frac{2}{\sqrt{3}} \frac{\rho}{c} \frac{N_{\theta}}{\Delta \theta_{s}} \frac{N_{\phi}}{2 \varphi_{t}}\int_{\chi_{d, i}}^{\chi_{u, i}} d \chi^{\prime}\int_{\varphi_{d}}^{\varphi_{u}} \cos \varphi^{\prime} \ \Bigg[i\bigg(\frac{1}{\gamma^{2}}+\varphi^{\prime 2} \notag \\ &+\chi^{\prime 2}\bigg) K_{\frac{2}{3}}(\xi) + \chi^{\prime}\bigg(\frac{1}{\gamma^{2}}+\varphi^{\prime 2}+\chi^{\prime 2}\bigg)^{1 / 2} K_{\frac{1}{3}}(\xi) \Bigg] d \varphi^{\prime}, \label{eq2}
	\end{align}
	\begin{align}
	& A_{\perp, i} \simeq \frac{2}{\sqrt{3}} \frac{\rho}{c} \frac{N_{\theta}}{\Delta \theta_{s}} \frac{N_{\phi}}{2 \varphi_{t}}\notag \\
	& \times\int_{\chi_{d, i}}^{\chi_{u, i}} d \chi^{\prime} \int_{\varphi_{d}}^{\varphi_{u}}\left(\frac{1}{\gamma^{2}}+\varphi^{\prime 2}+\chi^{\prime 2}\right)^{1 / 2}  K_{\frac{1}{3}}(\xi) \varphi^{\prime} \cos \varphi^{\prime} d \varphi^{\prime}, \notag \\\label{eq3}
	\end{align}
	where $\varphi_t$ is the opening angle of the bunch, $\theta_{s}$ is the angle of the footpoint for field line at the stellar surface, $\rho$ is the curvature radius, $\gamma$ is the Lorentz factor of bunch, $\xi=\omega \rho\left(1/\gamma^{2}+\varphi^{\prime 2}+\chi^{\prime 2}\right)^{3 / 2}/3c$, and $K_{\nu}(\xi)$ is the modified Bessel function, $\chi_{d, i}$ and $\chi_{u, i}$ are the lower and upper boundary of $\chi$, $\varphi_{d}$ and $\varphi_{u}$ are the lower and upper boundary of $\varphi$, respectively.  $A_{\|}$ and $ A_{\perp}$ are the polarized components of the amplitude along $\boldsymbol{\epsilon}_{\|}$ and $ \boldsymbol{\epsilon}_{\perp}$, where  $\boldsymbol{\epsilon}_{\|}$ is the unit vector pointing to the center of the instantaneous circle, and $\boldsymbol{\epsilon}_{\perp}=\boldsymbol{n} \times \boldsymbol{\epsilon}_{\|}$ is defined.  One can derive the Stokes parameters via the polarization amplitudes.
	
	\begin{figure}
		\centering
		\includegraphics[width=0.48\textwidth]{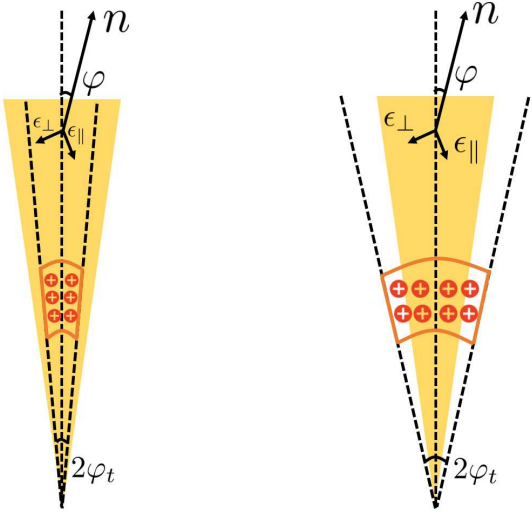}
		\caption{Schematic diagram of bunches seen in the horizon plane. The unit vector of the LOS is denoted by $\boldsymbol{n} $, and $\boldsymbol{\epsilon}_{\|}$ and $ \boldsymbol{\epsilon}_{\perp}$ denote the two polarization components. The opening angle of the emission region is 2$\varphi_t$. The yellow shaded area denotes emission cones with $1/\gamma$.  In the left panel, $0<\varphi_t\lesssim1/\gamma$. In the right panel, $1/\gamma\lesssim\varphi_t<10/\gamma$.}
		\label{Fig1}
	\end{figure}

\begin{figure}
		\centering
		\includegraphics[width=0.48\textwidth]{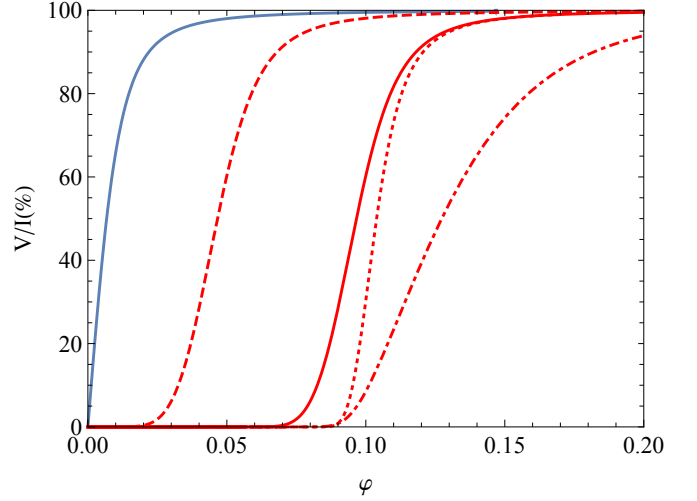}
		\caption{The CP fraction as a function of the angle between LOS and the trajectory plane. The blue line denotes the single particle and three-dimensional bunch case for $\varphi_t \lesssim 1/\gamma$, where $\gamma = 100$ and $\omega = \omega_c$. The red lines indicate three-dimensional bunch case for $\varphi_t>1/\gamma$, where dashed, solid, dot-dashed, and dotted lines denote $\gamma =100$, $\varphi_t=5/\gamma$, $\omega = \omega_c$; $\gamma =100$, $\varphi_t=10/\gamma$, $\omega = \omega_c$; $\gamma =10$, $\varphi_t=10/\gamma$, $\omega = \omega_c$; and $\gamma =100$, $\varphi_t=10/\gamma$, $\omega = 10\omega_c$, respectively. The other parameters are adopted as $\Delta \theta_{s} = 10^{-3}$ and $\rho = 10^7$ cm.}
		\label{Fig2}
	\end{figure}

	As shown in Figure \ref{Fig1}, we consider the two cases for $\varphi_t$ ($0<\varphi_t<1/\gamma$ and $1/\gamma<\varphi_t<10/\gamma$), and further study their polarization properties as follows. Figure \ref{Fig2} shows the CP as a function of the angle between the LOS and the trajectory plane ($\varphi$).  The CP approaches slowly 100$\%$ as $\varphi$ increases with a fixed $\omega$, $\gamma$, and $\varphi_t$. A single charge case is different from the three-dimensional bunche case for $\varphi_t > 1/\gamma$. However, a single charge shares similar polarization properties with the bunch case of $\varphi_t\lesssim1/\gamma$. Emitted radio waves have high LP if the LOS is limited to the beam within an angle of $1/\gamma$. In addition, the CP degree becomes stronger as $\varphi$ grows, due to the non-axisymmetric summation of $A_{\perp}$.
	
	Within the framework of coherent curvature radiation by bunches, $\varphi_t$ or bunch geometry can significantly affect the emission polarization. As shown in Figure \ref{Fig2}, if $\varphi_t$ becomes larger or $\gamma$ becomes smaller, the red solid and dot-dashed lines will move in the direction of increasing $\varphi$. Therefore, we conclude that the less event rate for higher observed CP suggests that  
	most highly linearly polarized bursts are generated within the emission cone. Besides, as $\omega$ becomes larger, we find the sharp evolution of CP with $\varphi$. The CP is difficult to be detected for the same LOS, since the $A_{\perp}$ decreases, leading to the original elliptical polarization is transformed into LP.

	\section{The Statistical Properties of Polarization}\label{sec3}
	\subsection{The CP-Flux Properties} \label{sec3.1}
	
\begin{figure}
		\centering
		\includegraphics[width=0.48\textwidth]{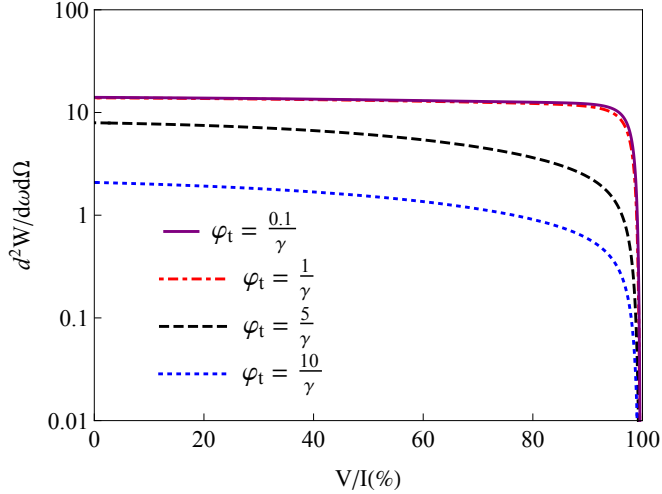}
		\caption{The flux as a function of CP for three-dimensional bunches. The purple solid line, red dot-dashed line, black dashed line, and blue dotted line indicate $\varphi_t=0.1/\gamma$, $1/\gamma$, $5/\gamma$, and $ 10/\gamma$, respectively. We adopt below physical parameters: $\gamma$ =100, $\rho=10^7$ cm, $\omega = \omega_c$,  $\Delta\theta_s = 10^{-3}$.}
		\label{Fig3}
	\end{figure}

	Within the framework of coherent curvature radiation by charged bunches, the CP is sensitive to the observed flux for the off-beam case. Some highly CP may appear at relatively low flux so that they generally are difficult to be detected by high sensitivity telescopes. In this section, we discuss the properties of CP-flux.
	According to Equations (\ref{eq1},\ref{eq2},\ref{eq3}), we can establish a relation between the flux and the CP.
	The flux as a function of CP for a three-dimensional bunch is shown in Figure \ref{Fig3}.
	We find that the flux is roughly constant and the CP can be observed within the opening angle of bunches. The upper limit of CP can be limited for different $\varphi_t$. However, the deviation of the LOS from the direction of the velocity makes the highly circularly polarized emission hard to be observed since the flux drops rapidly.  It is suggested that the flux is very sensitive for the off-beam case ($\varphi >\varphi_t$).

	\begin{figure}
		\centering
		\includegraphics[width=0.48\textwidth]{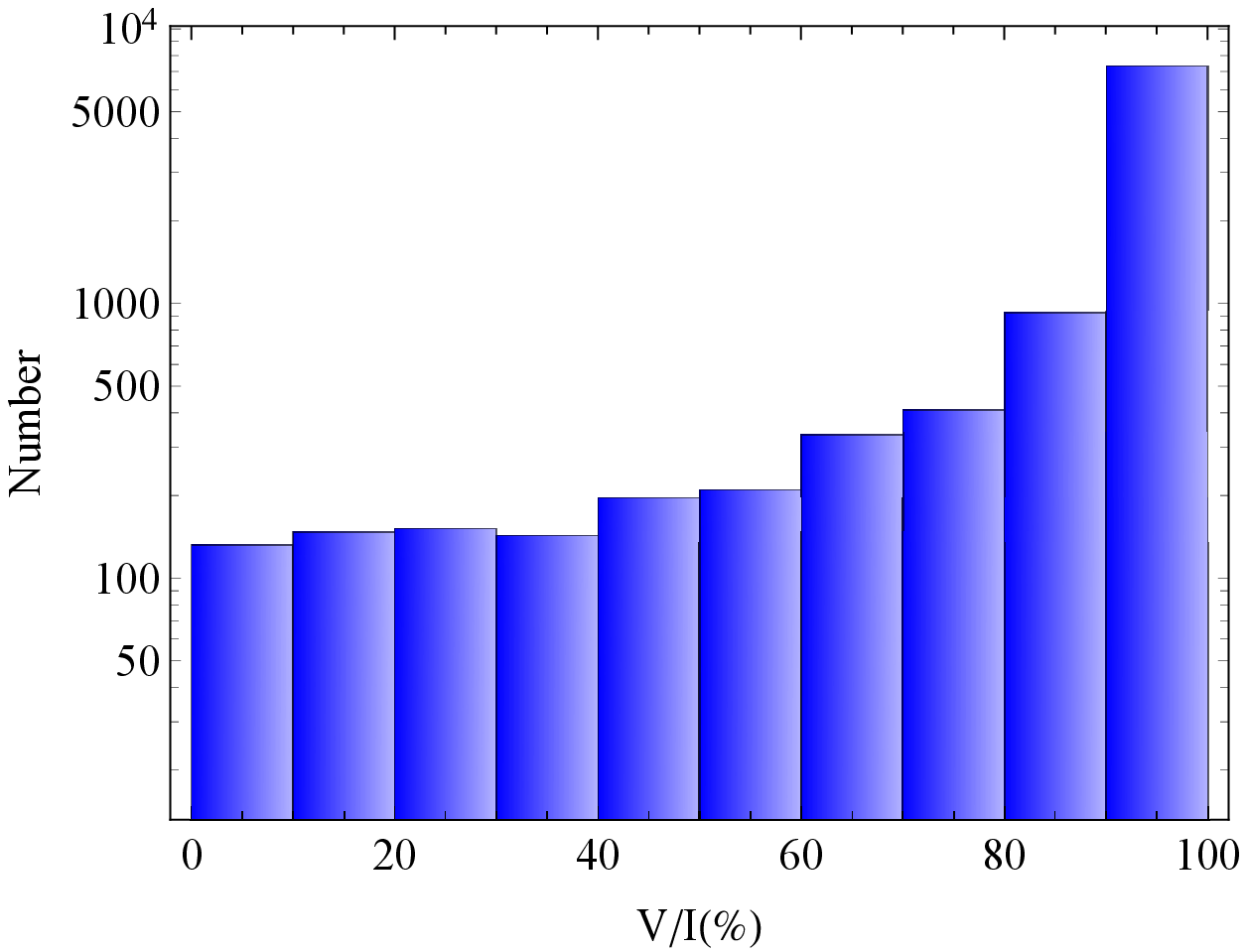}
		\includegraphics[width=0.48\textwidth]{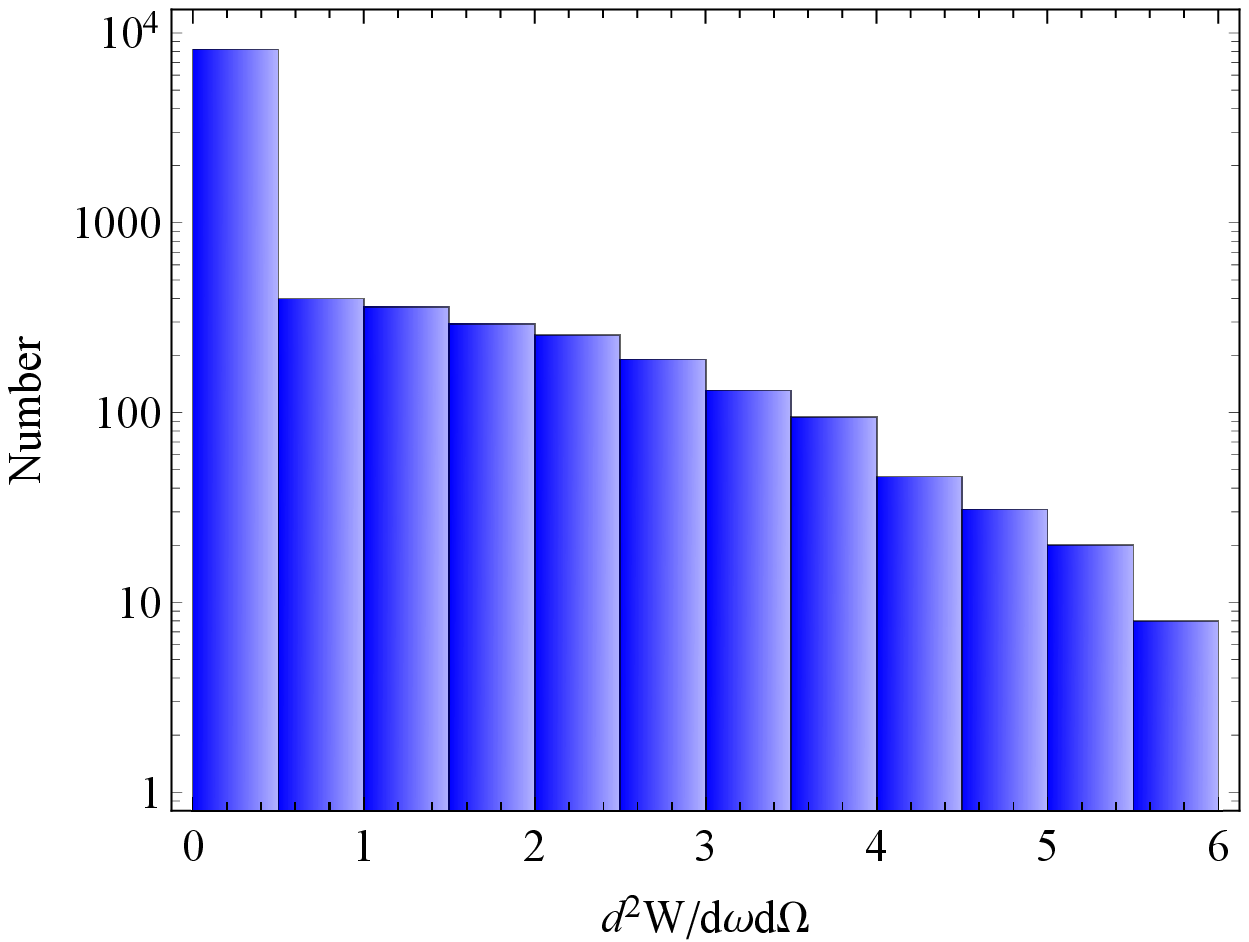}
		\includegraphics[width=0.48\textwidth]{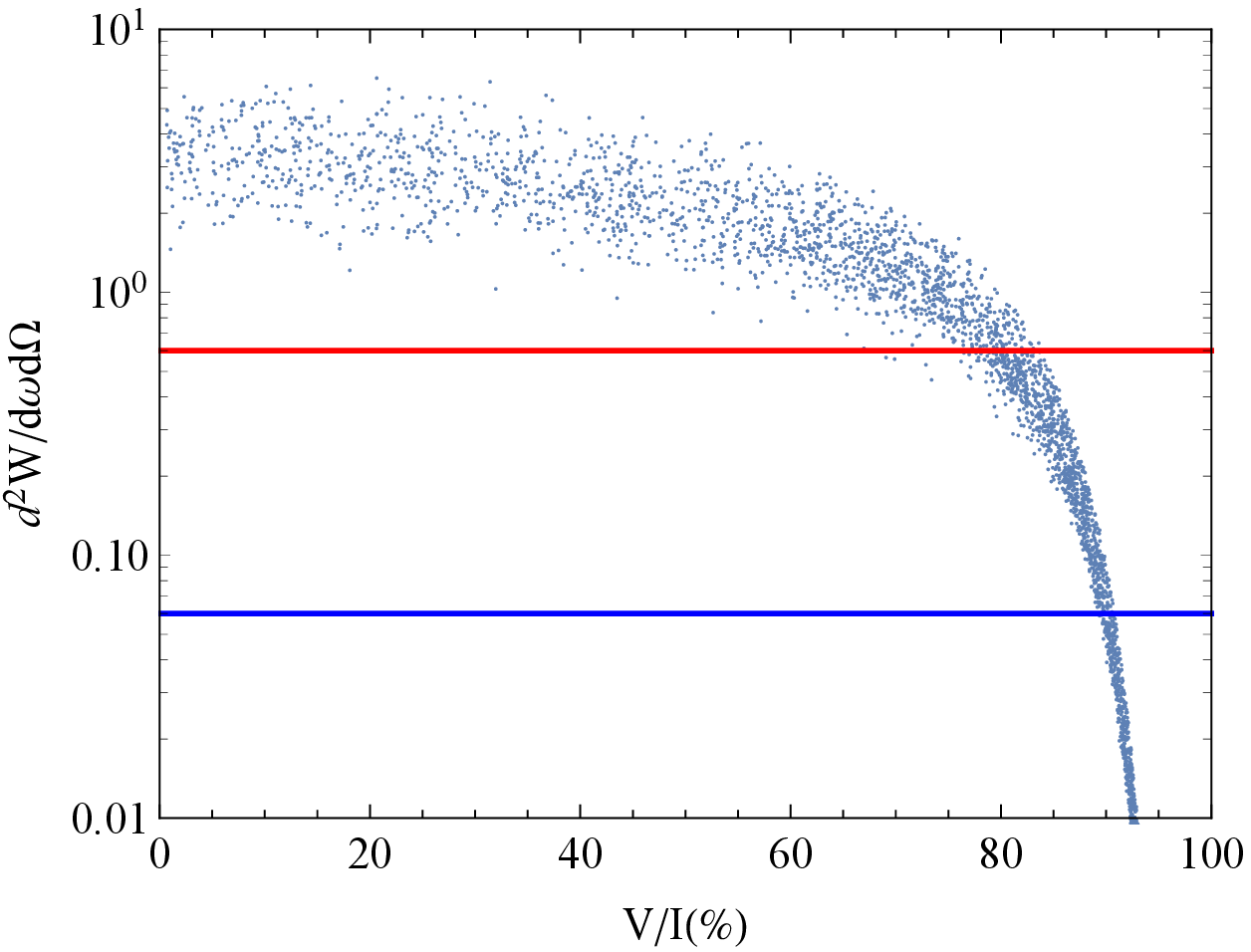}
		\caption{The upper panel, middle panel, and bottom panel denote the distribution of CP, the distribution of flux, and the scatter plot of flux-CP, respectively. We adopt below physical parameters: $\gamma$ = 100, $\omega = \omega_c$, $\rho=10^7$ cm, $\Delta\theta_s = 10^{-3}$. The unit of $d^2W/d\omega d\Omega$ is arbitrary. The red line and the blue line denote $10^{-1}$ and $10^{-2}$ times the maximum flux, respectively.  }
		\label{Fig4}
	\end{figure}

	We consider the angle $\varphi$ is randomly distributed in the range of $10^{-4}$ to $10^{-1}$, the LOS range (from $10^{-4}$ to $10^{-1}$) should be larger than the range of bunch opening angle (from $6\times10^{-3}$ to $1.4\times10^{-2}$). The angle $\varphi_t$ is a log-normal distribution in the range of $6\times10^{-3}$ to $1.4\times10^{-2}$ with a mean value of $1/\gamma$\footnote{In Figure \ref{Fig1}, we consider the two cases for $\varphi_t$ ($0<\varphi_t\lesssim1/\gamma$ and $1/\gamma\lesssim\varphi_t<10/\gamma$), $\varphi_t$ should be included in this range.}, and the  bunch length $L_{\rm b}$ is a log-normal distribution in the range of 10 to 50 cm.\footnote{The bunch length can be written as $L_{\rm b}=c/\pi\nu$, which affects the enhancement factor due to coherence. \citep{Yang2018,WangYang2021}.}. Assuming that the Lorentz factor, curvature radius, and angular frequency are constant ($\gamma=100,\rho=10^7 {\rm cm}, \omega=10^9 {\rm Hz}$), we scatter 10000 points above the range of the parameters. We then get the $A_{\perp}$ and $A_{\|}$ components, the CP, and flux can be calculated in the simulations.

	As shown in the upper panel of Figure \ref{Fig4}, we find that it has a more highly circularly polarized emission since more off-beam cases can appear. Besides, the smaller bunch opening angle trends appear in the off-beam case if the LOS is fixed.  It can be seen from the simulation result (see the bottom panel of  Figure \ref{Fig4}) that the CP can be constrained with $\lesssim 80\%$ within an order of magnitude of the maximum flux of FRB (the red line).
	The maximum flux of an FRB in our simulation is defined by the condition of $\varphi \simeq 10^{-4}$. The simulation of flux-CP shows that most of the bursts with high CP have relatively low flux.  The burst energies spans for more than two orders of magnitude, and an artificial sharp cutoffs in the energy distributions due to the instrumental limitation \citep{jiang2022}. We consider the flux threshold of FRBs $10^{-2}$ times less than the maximum flux (the blue line). Therefore, bursts with highly CP are hardly observed since the flux is too low.

\begin{figure}
		\centering
		\includegraphics[width=0.48\textwidth]{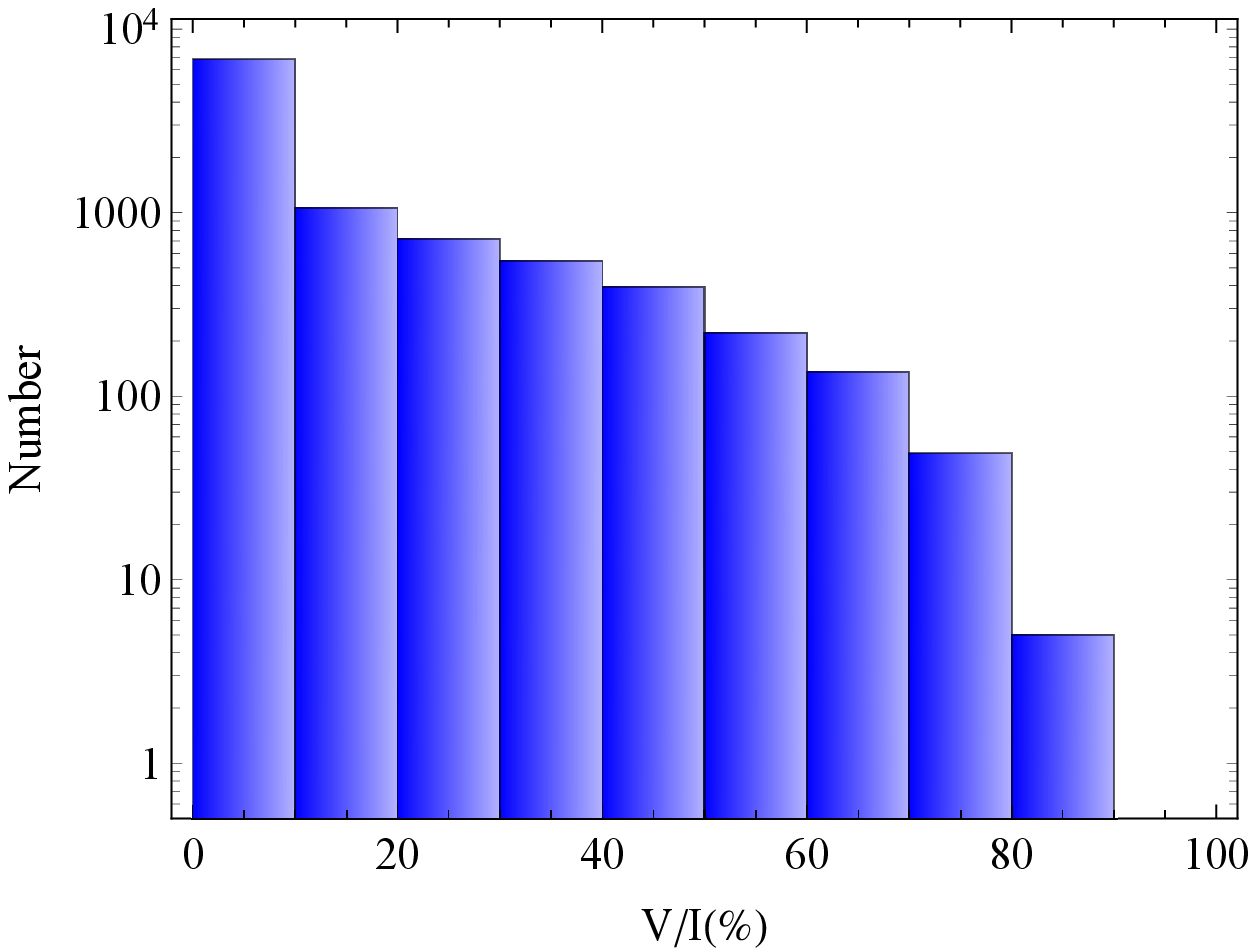}
		\includegraphics[width=0.48\textwidth]{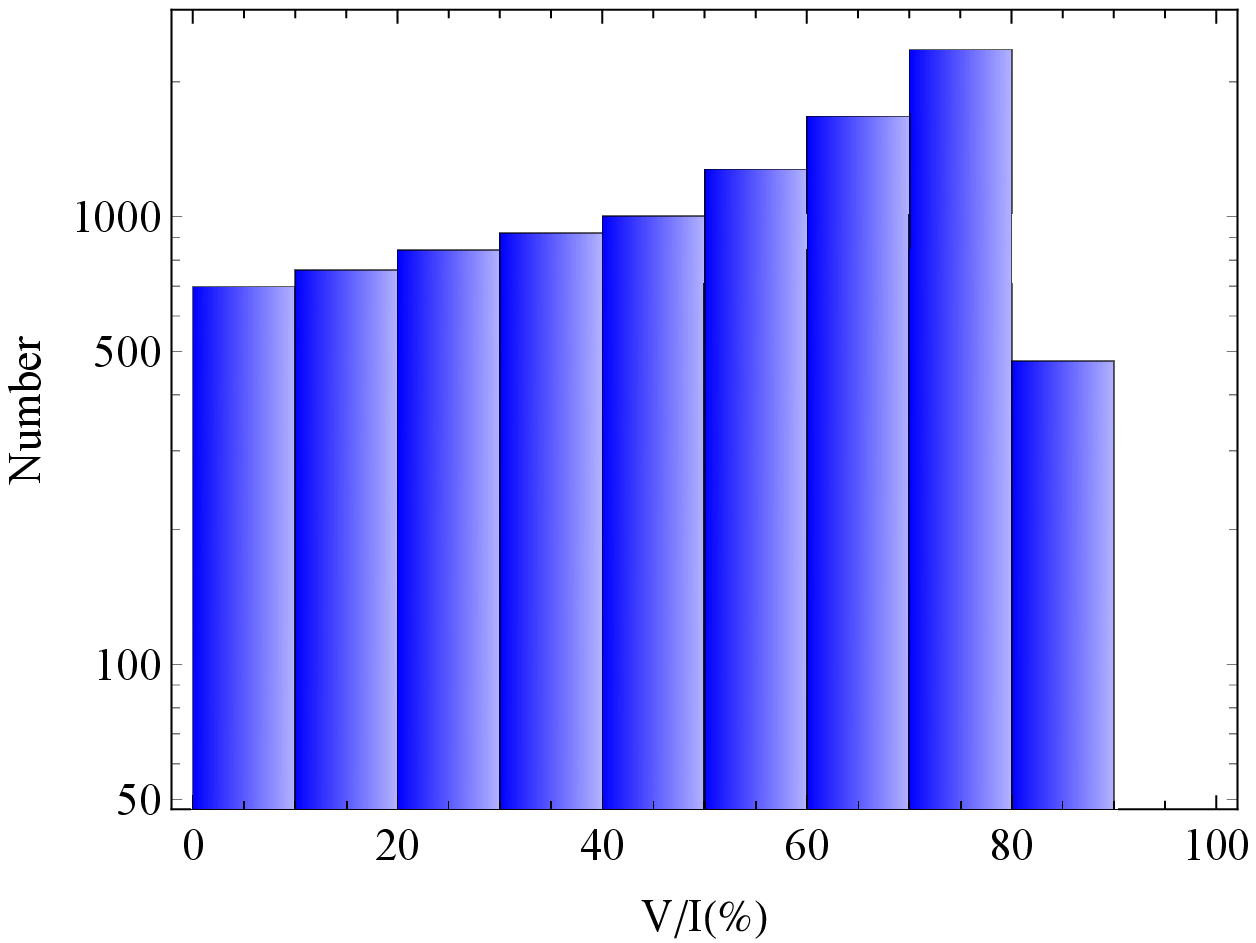}
		\caption{The distribution of CP for different $\varphi$ and $\varphi_t$. The upper panel denotes $\varphi_t = 5/\gamma$ and $\varphi$ varies from $10^{-4}$ to $10^{-1}$. The bottom panel denotes $\varphi_t = 1/\gamma$ and $\varphi$ varies from $10^{-4}$ to $10^{-2}$. We adopt below physical parameters:  $\gamma$ = 100, $\omega = \omega_c$, $\rho=10^7$ cm, $\Delta\theta_s = 10^{-3}$.}
		\label{Fig5}
	\end{figure}

	Some bursts with significant CP fraction can be as high as 75$\%$ \citep{XuH2021}, and a small fraction of the bursts have CP fraction higher than $70\%$ \citep{jiang2022}. It is suggested that the bunch opening angle is smaller for a fixed LOS or the off-beam case.  As shown in Figure \ref{Fig9}, we plot the observed range of circular polarization (CP) fractions and linear polarization (LP) fractions for repeating FRBs (also see the statistical data from Table 1 by \cite{WangJL2022} ). 
		The CP fractions for most repeaters are $\lesssim 15\%$. Besides, most of the bursts with low CP for FRB 20201124A \citep{jiang2022}  indicate that most bunches have large opening angles to contribute to the low CP distribution (see the upper panel of Figure \ref{Fig5} ).   Furthermore, the total degree of polarization is higher than 90$\%$ for most of the bursts of FRB 20201124A \citep{jiang2022}. It is infered that the depolarization caused by the propagation effects basically can be neglected.  Our simulation results are consistent with the representing observations.

	As shown in Figure \ref{Fig5}, we also simulate the distribution of CP for different $\varphi$ and $\varphi_t$. The upper panel of Figure \ref{Fig5} shows $\varphi_t = 5/\gamma$ and $\varphi$ varies from $10^{-4}$ to $10^{-1}$. 
	In our simulated results, we find $\sim 0.54\%$ of bursts have CP fraction higher than $70\%$ and  $\sim 80\%$ of bursts have CP fraction smaller than $20\%$ for $\varphi_t = 5/\gamma$ and $\varphi$ varies from $10^{-4}$ to $10^{-1}$, which is consistent with the observations of \cite{jiang2022}. 
	Compared with Figure \ref{Fig4}, the bunch opening angle $\varphi_t$ is larger for the same LOS range or the upper limit of $\varphi$ is smaller for the same $\varphi_t$ (the bottom panel of Figure \ref{Fig5}), making it has a more low CP distribution.
	It is consistent with the observations of \cite{jiang2022}. 
	Since the larger bunch opening angle for the same LOS range, most low circularly polarized bursts are generated within the emission cone.

	\subsection{Depolarization for FRBs} \label{sec3.2}
	
	Very recently, \cite{FY2022} reported the polarization measurements of five active repeaters and found a trend of higher depolarization at lower frequencies. It can be well explained by the multipath propagation through a magnetized inhomogeneous plasma screen, and consistent with the observed temporal scattering \citep{Yang2022}. In this section, we consider the frequency-dependent LP fraction within the framework of coherent curvature radiation and simulate the distribution of depolarization degree.

\begin{figure}
		\centering
		\includegraphics[width=0.48\textwidth]{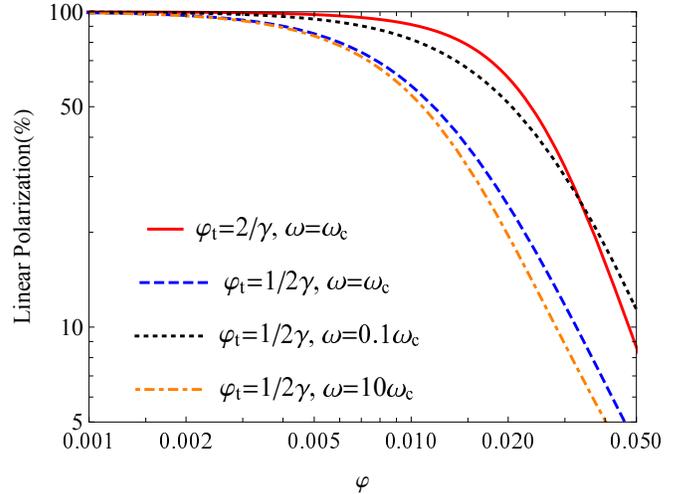}
		\caption{The degree of LP as a function
			of $\varphi$. The red solid line, blue dashed line, black dotted line, and orange dot-dashed line indicate $\varphi_t= 2/\gamma$, $\omega = \omega_c$; $\varphi_t=1/2\gamma$, $\omega = \omega_c$; $\varphi_t=1/2\gamma$, $\omega = 0.1\omega_c$; and $\varphi_t=1/2\gamma$, $\omega = 10\omega_c$, respectively. We adopt below physical parameters: $\rho=10^7$ cm, $\gamma=100$, $\Delta\theta_s = 10^{-3}$.}
		
		\label{Fig6}
	\end{figure}

	Polarized waves are generated by curvature
	radiation by relativistic particles streaming along curved magnetic field lines, which can be divided into the $A_{\perp}$ and $A_{\|}$ components. The polarized waves is nearly 100$\%$ LP for $A_{\perp} \sim 0$. As shown in Figure \ref{Fig6}, the LP decreases as the $\varphi$ increases with a fixed $\omega$ and $\varphi_t$, which is attributed to the $A_{\perp}$ contribution.  As the bunch opening angle $\varphi_t$ becomes smaller (see the red and blue line of Figure \ref{Fig6}), it is easier to appear the off-beam case, leading to the circularly polarized degree becomes strong. As shown in Figure \ref{Fig6} (see the blue, black, and orange lines), the depolarization degree has a small variation in a wide frequency band. 
	We find the LP fraction decreases at higher frequencies within the framework of coherent curvature radiation. The result is very similar to the frequency dependence of pulsar LP \citep{Morris1970,Manchester1971}. According to the radius-to-frequency mapping, high-frequency emission is generated from the lower magnetosphere, where the rotation effect gets weaker and the distribution regions of the two components (O-mode and X-mode) are overlapped \citep{WPF2015}. Thus, more significant depolarization would be observed for emission at a higher frequency. If the depolarization is caused by the propagation effect at the magnetosphere, a trend of lower LP at higher frequencies may be predicted.

\begin{figure}
		\centering
		\includegraphics[width=0.48\textwidth]{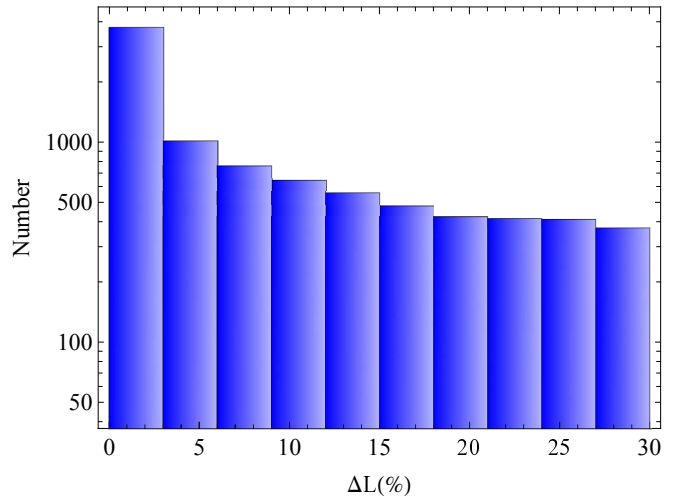}
		\caption{The simulated depolarization degree distribution. The frequency varies from 100 MHz to 10 GHz. We adopt below physical parameters: $\gamma$ =100, $\rho=10^7$ cm, $\omega = \omega_c$,  $\Delta\theta_s = 10^{-3}$.}
		\label{Fig7}
	\end{figure}

	We simulate the distribution of depolarization degree within the framework of coherent curvature radiation and consider the angle $\varphi$ is randomly distributed in the range of $10^{-4}$ to $10^{-1}$, and the angle $\varphi_t$ is a log-normal distribution in the range of $6\times10^{-3}$ to $1.4\times10^{-2}$ with a mean value of $1/\gamma$, and the frequency varies from 100 MHz to 10 GHz. The depolarization variation is shown in Figure \ref{Fig7}, most of the depolarization degrees of bursts have a small variation in a wide frequency band. We also constrain the depolarization degree variation within 30$\%$ for the frequency varies from 100 MHz to 10 GHz. \cite{FY2022} found the depolarization degree varies rapidly as the frequency for FRB sources. The larger depolarization infers that the rotation-measure scattering is large enough. The LP decreases by $\sim 100\%$ to low frequency for the FRBs with the frequency of half order of magnitude \citep{FY2022}. However, Figure \ref{Fig7} shows that the depolarization varies within  $\sim 30\%$ with the frequency of two orders of magnitude. Therefore, the contribution of the intrinsic curvature emission mechanism for the depolarization is very less than the rotation-measure scattering. It is consistent with the observation reported by \cite{FY2022}.
	
	\subsection{Simulated Polarization Angle Evolution} \label{sec3.3}

	An ‘S’ or ‘inverse S’ shape pattern generally can be observed in the radio pulsars \citep{Lorimer2012}. The profiles have been predicted by the rotating vector model \citep{RVM1969}.  Some FRBs have a flat polarization angle across each pulse, which may be caused by the emission from the outer magnetosphere or a slowly rotating  pulsar. However, emission from the inner magnetosphere or a quickly rotating  pulsar could explain the rapid swing of polarization angle across each pulse for some FRBs. Besides, the spin and magnetic inclination of neutron star also affect the PA evolution. In this section, we mainly simulate the PA evolution. The  polarization angle $ \Psi$ can be written as
	\begin{equation}
	\Psi=\frac{1}{2} \tan ^{-1}\left(\frac{U_{s}}{Q_{s}}\right),
	\end{equation}
	where  \begin{equation}
	\left(\begin{array}{c}U_{s} \\ Q_{s}\end{array}\right)=\left(\begin{array}{cc}\cos 2 \psi & \sin 2 \psi \\ -\sin 2 \psi & \cos 2 \psi\end{array}\right)\left(\begin{array}{l}U \\ Q\end{array}\right).
	\end{equation}
	$U$ and $Q$ are the Stokes parameters.  $ \psi$ as a function of azimuthal angle with respect to the spin axis $\phi$ \citep{RVM1969}.
	\begin{equation}
	\tan \psi=\frac{\sin \alpha \sin \phi}{\cos \alpha \sin \zeta-\cos \zeta \sin \alpha \cos \phi},
	\end{equation}    
	where $\alpha$ is the angle between the magnetic axis and the rotational axis, and $\zeta$ is the angle between the LOS and spin axis.
	
	\begin{figure}
		\centering
		\includegraphics[width=0.48\textwidth]{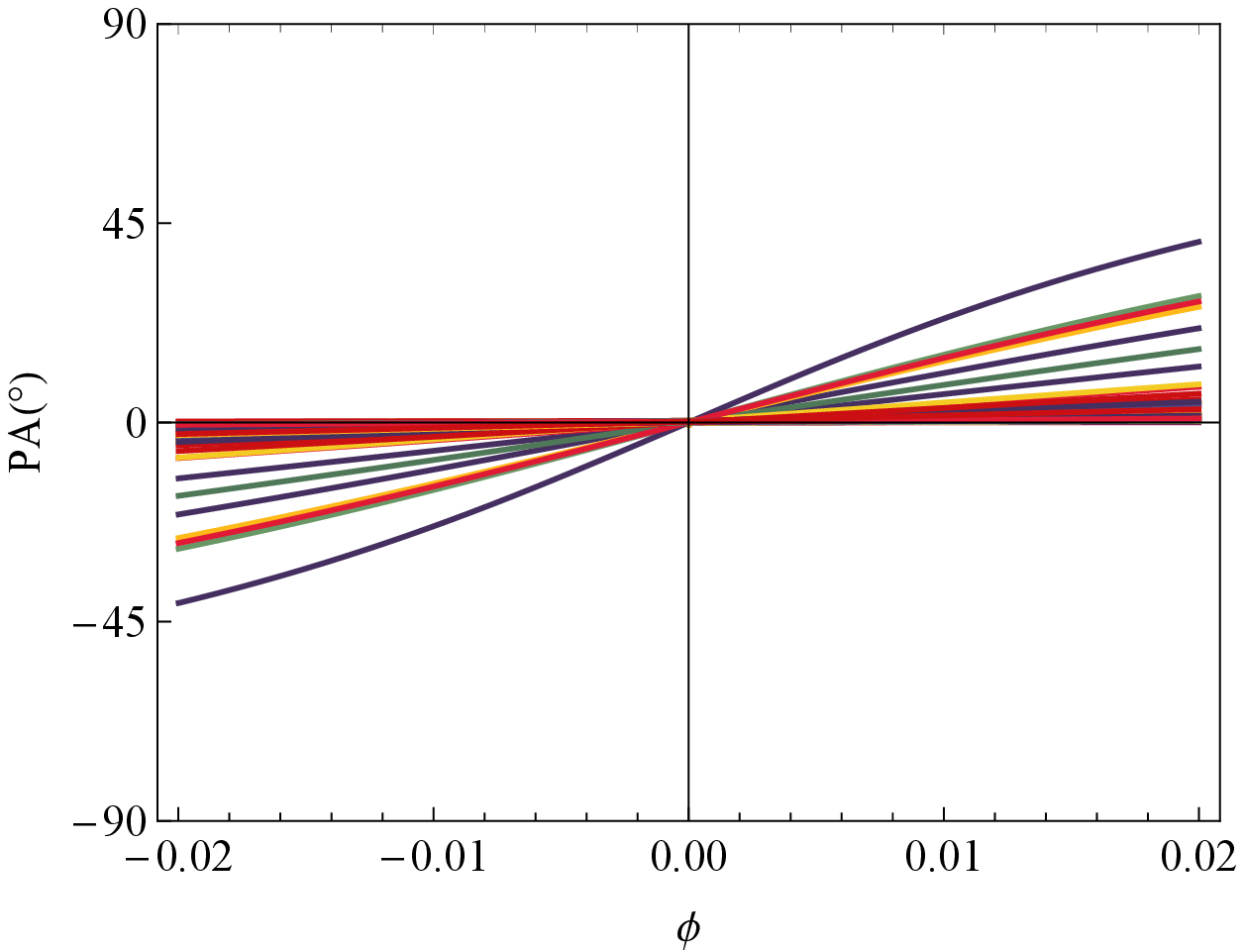}
		\includegraphics[width=0.48\textwidth]{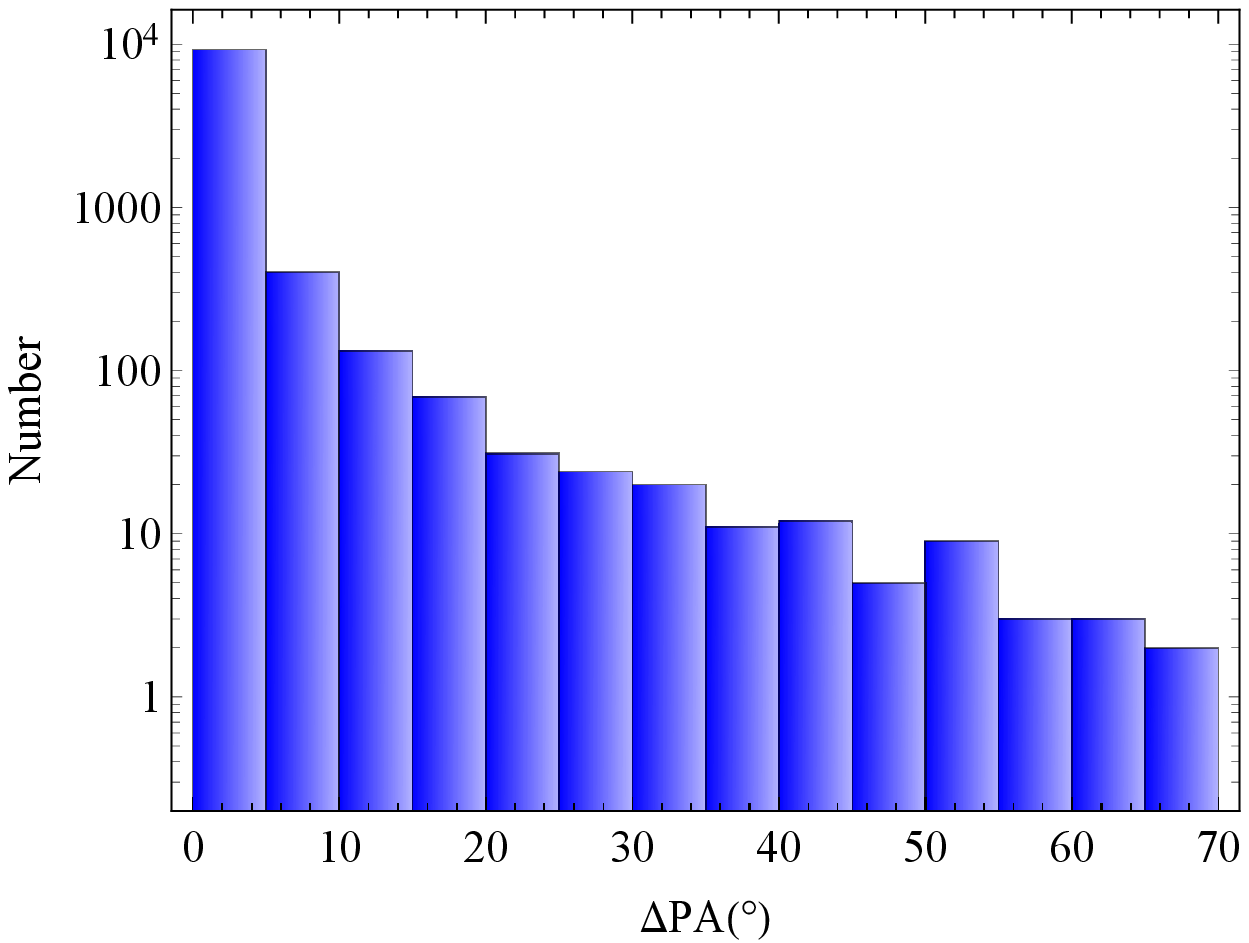}
		\caption{Simulated polarization angle evolution and variation for each pulse. The upper panel denotes the simulated PA evolution. The different lines denote the flatter PA evolution within the burst phases for $\varphi<\varphi_t$. The bottom panel denotes simulated PA variation.  The parameters are adopted as follows: $\alpha=\pi / 6,  \zeta=\pi / 4$, $\gamma$ =100, $\rho=10^7$ cm.}
		\label{Fig8}
	\end{figure}
	
\begin{figure*}
		\centering
		\includegraphics[width=0.9\textwidth]{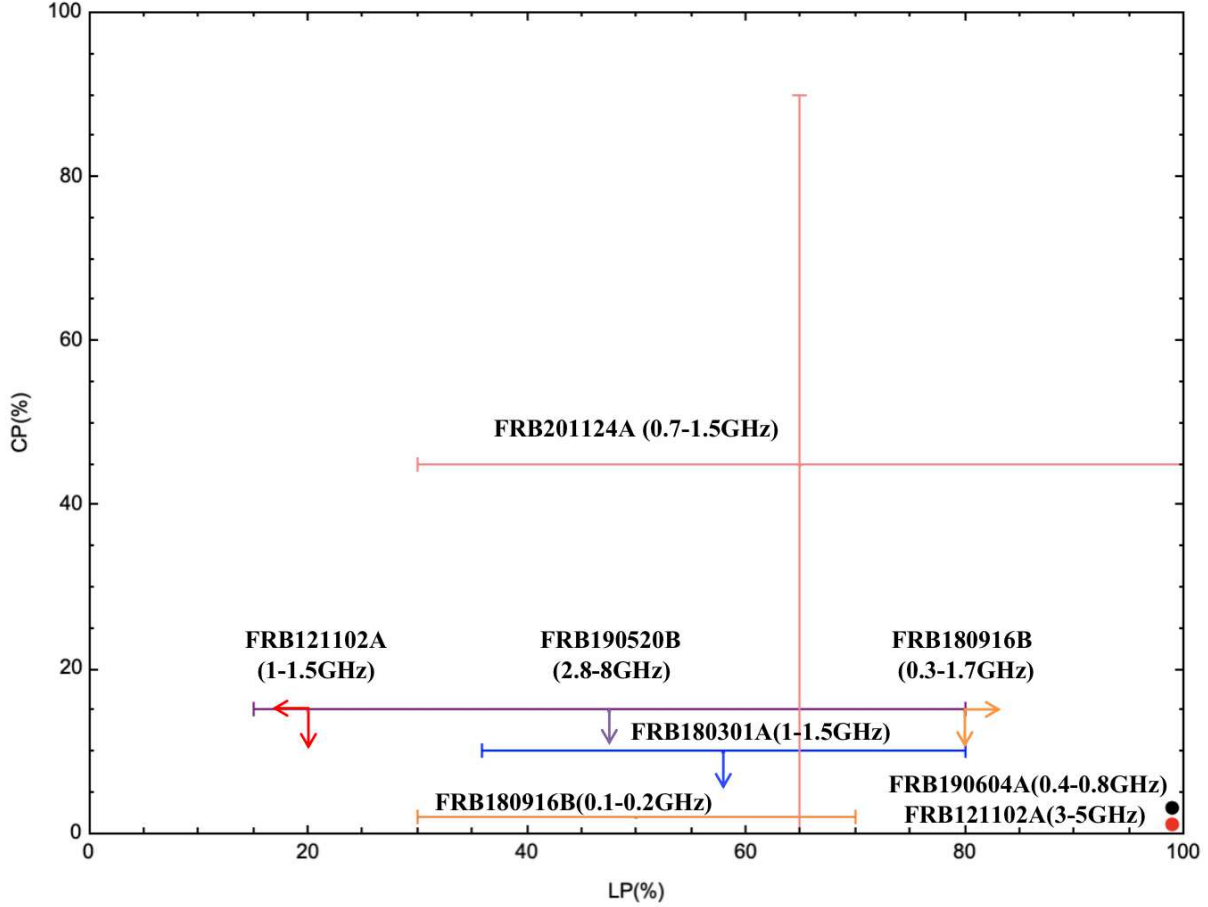}
		\caption{The observed ranges of circular polarization (CP) fractions and linear polarization (LP) fractions for repeating FRBs. The red point: FRB 121102A   \citep[3-5 GHz, PA constant,][]{Michilli2018,HilmarssonM2021}; the red arrows: FRB 121102A  \citep[1-1.5 GHz, PA constant,][]{Plavin2022}; the blue arrow: FRB 180301A \citep[1-1.5 GHz, PA constant or varying,][]{Luo2020}; the orange arrows: FRB 180916B \citep[0.3-1.7 GHz, PA constant,][]{Nimmo2021,Sand2022}; the orange line segment: FRB 180916B \citep[0.1-0.2 GHz, PA constant,][]{Pleunis2021}; the purple arrow: FRB 190520B \citep[2.8-8 GHz, PA constant,][]{AnnaThomas2022,DS2022,Niu2022}; the black point: FRB 190604A \citep[0.4-0.8 GHz, PA constant,][]{Fonseca2020}; the pink line segments: FRB 201124A \citep[0.7-1.5 GHz, PA constant or varying, ][]{jiang2022,Hilmarsson2021,Kumar2021,XuH2021}. The arrows denote the upper/lower limits of observations and the line segments denote the observed ranges of CP and LP. }
		\label{Fig9}
	\end{figure*}

	As shown in Figure \ref{Fig8}, we simulate the PA evolution across each pulse and PA variation distribution. We consider the angle $\varphi$ is randomly distributed in the range of $10^{-4}$ to $10^{-1}$, and the angle $\varphi_t$ is a log-normal distribution in the range of $6\times10^{-3}$ to $1.4\times10^{-2}$) with a mean value of $1/\gamma$ for each pulse. Most bursts show a flat PA evolution within the burst phases, some bursts present a swing of PA across each pulse. A flat PA evolution for $\varphi<\varphi_t$ and the PA evolution is more dramatic for $\varphi>\varphi_t$. We define the difference between maximum and minimum values of PA for each pulse as PA variation.  We find the maximum change in the PA across the pulse profile is less than $10^{\circ}$ for most of the bursts, and a small fraction of bursts have the PA variation across each pulse higher than $30^{\circ}$. It is suggested that most bursts have on-beam case to produce the small PA variation (e.g., FRB 20190520B, FRB 20121102A, FRB 20180301A, etc). Our simulated results are consistent with the predicted by \cite{WangYang2021} and the observations \citep{jiang2022,DS2021,Nimmo2021,CHIME2019,Gajjar2018,Michilli2018,Luo2020,Cho2020}. As shown in Figure \ref{Fig2}, most of low CP bursts are generated within the emission cone. Therefore, the small PA variation is correlated with the low CP in the on-beam case. Our conclusions are consistent with the observations (see Figure \ref{Fig9}).

	\section{ Conclusions and Discussion}\label{sec4}
	
	In this paper, we investigated the polarization features within the framework of coherent curvature radiation by charged bunches in the neutron star magnetosphere. We considered that FRBs are produced by coherent curvature radiation from bunches in the neutron star magnetosphere and discussed the statistical properties of polarization and the simulation results for radio bursts from an FRB repeater. The following conclusions can be drawn:

	$\bullet$ The CP across burst approaches slowly 100$\%$ as $\varphi$ increases with a fixed $\omega$,$\gamma$, and $\varphi_t$. A single charge case is different from the three-dimensional bunche case for $\varphi_t > 1/\gamma$. However, the two cases are equivalent for $\varphi_t \lesssim 1/\gamma$, since the opening angle of the bunch can be seen as a point source case. Emitted radio waves have high LP if the LOS is limited to the beam within an angle of $1/\gamma$. While the CP degree becomes stronger as $\varphi$ grows, due to the non-axisymmetric summation of $A_{\perp}$. Within the framework of coherent curvature radiation, the bunch opening angle $\varphi_t$ will significantly affect the polarization of emission. The less event rate for observed high CP suggests that most highly linearly polarized bursts are generated within the emission cone.

	$\bullet$ The flux is almost constant within the opening angle of bunches. However, when the LOS derivates from the bunch opening angle, the larger the derivation, the larger the CP but the lower the flux. We simulated the distribution of CP-flux, and found that most of the bursts with high CP have relatively low flux, and $\sim 0.54\%$ of bursts have CP fraction higher than $70\%$ and $\sim 80\%$ of bursts have CP fraction smaller than $20\%$ for $\varphi_t = 5/\gamma$ and $\varphi$ varies from $10^{-4}$ to $10^{-1}$. Besides, the CP can be constrained with $\lesssim 80\%$ within an order of magnitude of the maximum flux of FRB  for $\varphi_t = 1/\gamma$.

	$\bullet$ Most of the depolarization degrees of bursts have a small variation in a wide frequency band since the slow evolution of LP as frequency. We also constrained the depolarization variation within 30$\%$ for the frequency varies from 100 MHz to 10 GHz. Therefore, the contribution of the intrinsic curvature emission mechanism for the depolarization is very less than the rotation-measure scattering. Furthermore, we simulated the PA evolution and found that the maximum change in the PA across the pulse profile is less than $10^{\circ}$ for most of the bursts, and a small fraction of bursts have the PA variation across each pulse higher than $30^{\circ}$. It is suggested that most bursts have on-beam case to produce the small PA variation.  In conclusion, our simulation results are consistent with the representing observations \citep{jiang2022}.

	A significant CP was observed in an FRB repeater, FRB 20201124A \citep{jiang2022,Hilmarsson2021,Kumar2021,XuH2021}. 
	We considered that the high CP fraction may be caused by a coherent curvature radiation mechanism \citep{WangJiang2021,WangYang2021,Tong2022}. As shown in Figure \ref{Fig2}, some important ingredients in the production of the CP are the geometric configuration of the bunch (e.g. the $\varphi_t$ decreases for the same LOS), and the perpendicular flux $A_{\perp}$ increases. Although the CP fraction can be up to 75$\%$ for some bursts, depolarization exists in the process of incoherent summation of CP of opposite senses. It would cause some unpolarized emission. In addition, the CP is canceled by incoherent summation for the asymmetry of sparking distribution \citep{Radhakrishnan1990,WangYang2021}.  Notably, \cite{Kumar2021} detected the CP of some bursts and the flux varies in a small range, the flux-CP shows a dispersion distribution. It is mainly caused by the  fluctuation (e.g. the bunch length, opening angle).

	According to the dipole magnetosphere geometry, the maximum cross section of the bunch can be approximately given by $\lambda R_{\rm b}$, where $\lambda$ is the wavelength of the bunch, and $R_{\rm b}$ is the radiation radius of the bunch. If there is a big phase gap between adjacent bunches, the emission is incoherent. The incoherent critical angle between adjacent magnetic lines is $\Delta\theta_{\rm b}=(\lambda/ {\rm R_b})^{1/2}$. Assuming $\lambda$ = 10 cm and $R_{\rm b} = 10^7$ cm, we find $\Delta\theta_{\rm b} = 10^{-3}$. The fractional reduction in the LP amplitude can be given by $f_{\mathrm{depol}} \equiv 1-\sin (\Delta \theta)/\Delta \theta$, where $\Delta \theta$ is the intra-channel polarization position angle rotation. According to the above estimation, we can obtain the LP is nearly 100$\%$, and the depolarization degree is nearly 0. This result is consistent with the polarization observation of repeating FRBs \citep{DS2021,Nimmo2021,CHIME2019,Gajjar2018,Michilli2018}. Besides, \cite{Liu2020} simulated the LP distribution for the repeating FRBs and constrain the LP with $\gtrsim 30\%$ for the FRBs with a flux of an order of magnitude lower than the maximum flux. However,  the depolarization is mainly caused by the large differences of multi-path rotation measures when a radio wave propagates in the magneto-ionic inhomogeneous environment \citep{FY2022}.
	
	As shown in Section \ref{sec3.2}, we have discussed the frequency-dependent linear polarization properties.  We here discuss the time-dependent linear polarization properties, which would help test radiation mechanisms.  The observed time-frequency is downward drifting from most repeating FRBs \citep{Gajjar2018,Michilli2018,Hessels2019,Josephy2019,Caleb2020,Day2020,Platts2021}. The time-frequency downward drifting is a natural consequence of coherent curvature radiation \citep{Wang2019}. A spark observed at an earlier time with a higher frequency is emitted in a more-curved magnetic field line.  As shown in Figure \ref{Fig6}, the 
	LP fraction decreases at higher frequencies within the framework of coherent curvature radiation. Combined with the time-frequency observation properties of FRBs, we find that a lower linear polarization fraction is observed at an earlier time with a higher frequency for the same $\varphi_t$, and the depolarization degree has a small variation within the observation time. The above time-dependent linear polarization properties can be used to test the radiation mechanism of FRBs.

	PA evolution generally involves two effects: the rotation radiation beam based on the rotating vector model \citep{RVM1969} and off-axis polarization properties of curvature radiation. According to the rotation vector model, the PA is flat for a slow rotating pulsar but evolves at near the smallest impact angle.
	The Stokes parameters $U$ and $Q$ of this accumulated emission also play a part in PA evolution.  A flat PA evolution (e.g. FRB 121102, FRB 180916, and FRB 20201124A) indicates a slow rotating pulsar and on-beam case. Variable PA evolution (FRB 180301, 181112) may need more complicated magnetic field configurations and LOS geometry.

	Motivated by an inverse Compton scattering (ICS) model of radio pulsar radiation \citep{Qiao1998,Xu2000}. Recently, \cite{ZB2022} proposed a model invoking coherent ICS by bunches as the radiation mechanism of FRBs. Considering the coherency of the radiation from a bunch of electrons, the radio radiation of pulsar from the lower emission altitudes would produce the CP \citep{Xu2000}. The polarization properties of the scattered photons will be affected by the asymmetrical particle distribution, the angular frequency of the scattered wave, and so on. We will further study the spectra and polarization features of the coherent ICS model in our future work.

	\acknowledgments
	
	We are grateful to Fa-Yin Wang and an anonymous referee for helpful discussions and constructive comments. This work was supported by the National
	Key Research and Development Program of China (grant
	Nos. 2017YFA0402600, 2018YFA0404204), the National SKA Program of China (grant No. 2020SKA0120300), the National Natural Science
	Foundation of China (grant No. 11833003, U2038105) and the Program for Innovative Talents, Entrepreneur in Jiangsu.
	W.-Y.W is supported by a Boya Fellowship and the fellowship of China Postdoctoral Science Foundation No. 2021M700247 and No. 2022T150018.
	Y.-P.Y is supported by National Natural Science Foundation of China (grant No. 12003028), the National Key Research and Development Program of China (2022SKA0130101) and the China Manned Spaced Project (CMS-CSST-2021-B11).

\end{document}